\documentclass[pra, aps, showpacs, reprint]{revtex4-1}

\usepackage[utf8]{inputenc}
\usepackage{graphicx}
\usepackage{amsmath}
\usepackage{amssymb}
\usepackage{braket}
\usepackage{color}
\usepackage{tikz}
\usetikzlibrary{positioning}
\usepackage{pdfpages}

\makeatletter
\AtBeginDocument{\let\LS@rot\@undefined}
\makeatother

\newcommand{\diff}[1]{\mathrm{d} #1}

\newcommand{\Imag}{{\mathrm{Im}}}
\newcommand{\Real}{{\mathrm{Re}}}
\newcommand{\ve}[1]{\boldsymbol{\mathbf{#1}}}

\providecommand{\abs}[1]{\lvert#1\rvert} 

\definecolor{darkgreen}{RGB}{0, 150, 0}
\definecolor{cyan2}{RGB}{0, 255, 255}

\begin{document}

\author{{\O}yvind Johansen}
\author{Hans Skarsv{\aa}g}
\author{Arne Brataas}
\affiliation{Center for Quantum Spintronics, Department of Physics, Norwegian University of Science and Technology, NO-7491 Trondheim, Norway}

\date{\today}
\title
{Spin-transfer Antiferromagnetic Resonance}

\begin{abstract}
Currents can induce spin excitations in antiferromagnets, even when they are insulating. We investigate how spin transfer can cause antiferromagnetic resonance in bilayers and trilayers that consist of one antiferromagnetic insulator and one or two metals. An ac voltage applied to the metal generates a spin Hall current that drives the magnetic moments in the antiferromagnet. We consider excitation of the macrospin mode and of transverse standing-spin-wave modes. By solving the Landau--Lifshitz--Gilbert equation in the antiferromagnetic insulator and the spin-diffusion equation in the normal metal, we derive analytical expressions for the spin-Hall-magnetoresistance and spin-pumping inverse-spin-Hall dc voltages. In bilayers, the two contributions compensate each other and cannot easily be distinguished. We present numerical results for a MnF$_2|$Pt bilayer. Trilayers facilitate separation of the spin-Hall-magnetoresistance and spin-pumping voltages, thereby revealing more information about the spin excitations. We also compute the decay of the pumped spin current through the antiferromagnetic layer as a function of frequency and the thickness of the antiferromagnetic layer. 
\end{abstract}

\maketitle
\section{Introduction}
Antiferromagnets have many qualities that make them attractive for use in spintronic devices. For example, the absence of stray fields allows for a dense storage of components without undesired crosstalk between the active elements. The most interesting feature of antiferromagnets is that their high resonance frequencies pave the way toward terahertz circuits \citep{Cheng:prl2016}.

Current-induced spin-transfer torques (STTs) can induce ferromagnetic resonance \cite{Liu:prl:2011} in both metallic and insulating ferromagnets \cite{Chiba:pra:2014,Schreier:prb:2015}. Antidamping-like STT is even under magnetization reversal \cite{Slonczewski1996}. Consequently, the magnetic moments in the two sublattices of a collinear antiferromagnetic insulator experience the same STT, which enables STT-driven spin dynamics in antiferromagnets. Current-induced STT is a powerful method for probing the magnetization dynamics in magnetic layers \cite{Sklenar:prb:2015,He:apl:2016}. An electric signal can simultaneously drive and detect the magnetization dynamics.  An ac voltage leads to an alternating spin current through the spin Hall effect \cite{Dyakonov1971}, which drives the magnetic moments at resonance. Subsequently, the spin Hall magnetoresistance (SMR) and spin pumping (SP) induce  dc voltages through the inverse spin Hall effect (ISHE) \cite{Saitoh:apl:2006} that can be measured using a bias tee.

SMR \cite{Nakayama:prl:2013,Chen:prb:2013} is the dependence of the normal metal resistance on the orientation of the magnetic moments in an adjacent magnetic layer relative to the applied current. When the magnetic moments precess, the resistance of the metal correspondingly oscillates. The mixing of the oscillating resistance and charge current generates a dc voltage bias that can provide insights into the magnetization dynamics. Recent experiments have indicated that SMR also occurs in antiferromagnetic insulator/normal metal (AF$|$N) bilayers \cite{Han:prb:2014,Hou:prl:2017,Hoogeboom2017,Baldrati2017,Fischer2017}. Theoretical predictions have also been made for conducting antiferromagnets \cite{Manchon:2017}. 

Similar to ferromagnets, SP is also active in antiferromagnets \cite{Cheng:prl2014}. However, to the best of our knowledge, there are no direct experimental detections of antiferromagnetic SP. The lack of direct experimental signatures is possibly due to the high resonance frequencies and low susceptibilities of the magnetic moments in antiferromagnets, which make experimental detection challenging. However, we have recently theoretically shown that the susceptibilities and thus the dc SP substantially increase near the spin-flop transition, where the resonance frequency is low \cite{Johansen:prb:2017}. Therefore, we expect that antiferromagnetic SP will be a prominent effect if we drive our system close to the spin-flop transition.

In this paper, we compute the dc voltages resulting from SMR and SP in antiferromagnetic insulator/normal metal bilayers. The driving source is an ac voltage bias on the normal metal. 
In addition to the macrospin mode, we also consider the excitation of transverse standing spin waves. These standing waves have a higher resonance frequency than that of the uniform precession modes, and these waves can be excited by tuning the frequency of the applied voltage bias.  The detection of such waves would reveal a wide variety of properties of the antiferromagnetic material.
The resonance frequencies can be used to determine contributions to the free energy of the antiferromagnet, such as exchange  and anisotropy frequencies, and the exchange lengths of the sublattices. 
The amplitudes and linewidths of the resonance peaks can also be used to determine both the intrinsic and SP-induced damping and thus the transverse spin conductance of the AF$|$N interface. Finally, we also show how the SMR and SP dc voltages can be separated by sandwiching the antiferromagnetic material between two metals and measuring the dc biases in the metals independently.
This approach requires that the dissipation of the pumped spin current through the antiferromagnet is negligible. 
We therefore study for what thicknesses of the antiferromagnetic layer and for what resonance frequencies this is a valid assumption.

\section{Model}
\begin{figure}[h!]
\centering
\begin{tikzpicture}
\node[above right] (img) at (0,0) {\includegraphics[width=\linewidth]{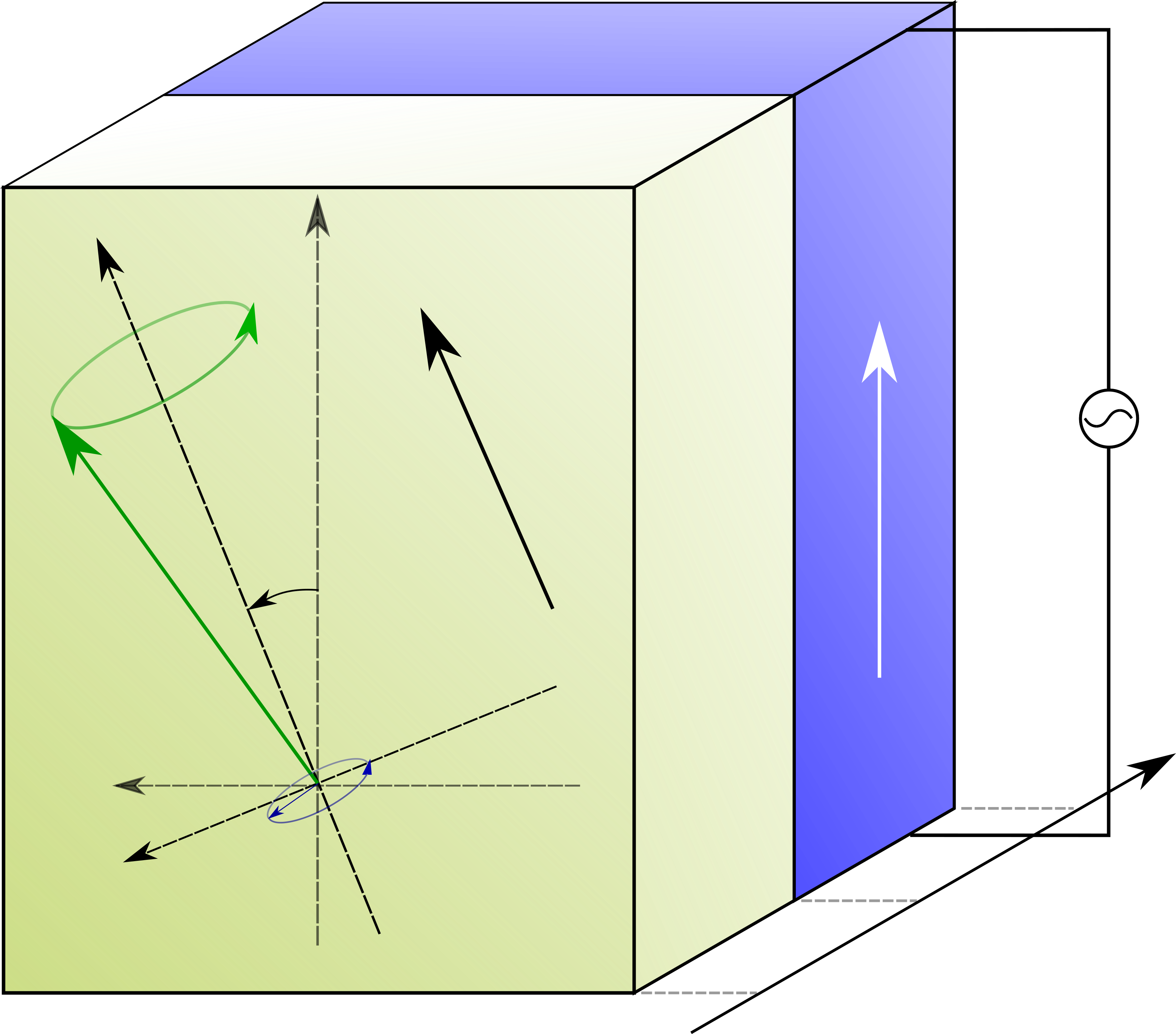}};
\node at (55pt,40pt) {\textcolor{blue}{\large{$\ve{m}(t)$}}};
\node at (18pt,103pt) {\textcolor{darkgreen}{\Large{$\ve{n}(t)$}}};
\node at (188pt,165pt) {\textcolor{white}{\Large{$\ve{I}_c(t)$}}};
\node at (218pt,132pt) {\textcolor{black}{\Large{$V_\text{ac}$}}};
\node at (62pt,105pt) {{\Large{$\theta$}}};
\node at (100pt,100pt) {{\Large{$\ve{H}_0$}}};
\node at (78pt,172pt) {{\Large{$x'$}}};
\node at (15pt,167pt) {{\Large{$x$}}};
\node at (22pt,55pt) {{\Large{$z'$}}};
\node at (20pt,35pt) {{\Large{$z$}}};
\node at (245pt,67pt) {{\large{$y$}}};
\node at (219pt,55pt) {{{$y=d_\text{N}$}}};
\node at (213pt,30pt) {{{$y=0$}}};
\node at (190pt,10pt) {{{$y=-d_\text{AF}$}}};
\node at (90pt,190pt) {\textcolor{black}{\huge{AF}}};
\node at (120pt,210pt) {\textcolor{white}{\huge{N}}};
\end{tikzpicture}
\caption{An ac voltage applied to a normal metal with strong spin-orbit coupling generates spin currents that flow into the antiferromagnetic insulator, exciting the magnetic moments. The direction of the applied voltage and the easy axis of the antiferromagnet are parallel to the AF$|$N interface, and there is an angle of $\theta$ between them. An external field $\ve{H}_0$ along the easy axis of the antiferromagnet controls the resonance frequency and the magnetic susceptibilities.}
\label{fig:Bilayer}
\end{figure}
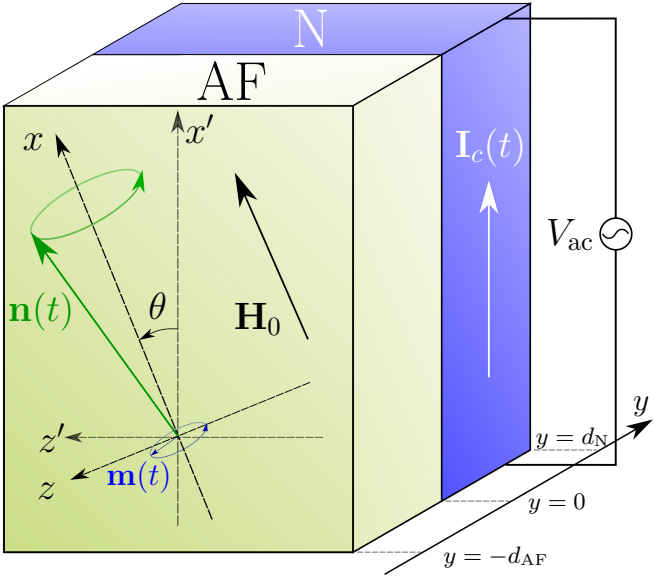

We first consider a bilayer that consists of an antiferromagnetic insulator in contact with a heavy metal, as shown in Fig.\ \ref{fig:Bilayer}. An ac voltage applied to the metal causes spin excitations in the antiferromagnet via the spin Hall effect and the STT. Subsequently, SP in combination with the ISHE and SMR cause a dc voltage in the normal metal. 

For the bilayer system, our main result is the analytical expressions for the dc voltages resulting from SP and SMR in Eq.\ \eqref{eq:VDC}. These expressions hold for a uniaxial antiferromagnet under the influence of an external magnetic field that can control the resonance frequency and enhance the voltage signals \cite{Johansen:prb:2017}. For the N$|$AF$|$N trilayer, our main contribution is illustrating how this system can be used to measure the SP and SMR voltages independently. These contributions cannot be distinguished in the bilayer system because they have the same frequency dependence. 

\subsection{Equations of motion}
The sublattice magnetizations of the antiferromagnetic insulator are $\ve{M}_1$ and $\ve{M}_2$. We describe the dynamics of these magnetizations in terms of the dimensionless average magnetization and N\'{e}el order parameter vectors $\ve{m}$ and $\ve{n}$, which are defined as $L \ve{m} = (\ve{M}_1+\ve{M}_2)/2$ and $L \ve{n} = (\ve{M}_1-\ve{M}_2)/2$, where $L$ is the saturation magnetization of each sublattice. These vectors satisfy the constraints $\ve{m}^2+\ve{n}^2=1$ and $\ve{m}\cdot\ve{n}=0$. The coupled equations of motion for $\ve{m}$ and $\ve{n}$ are given by the Landau--Lifshitz--Gilbert (LLG) equations 
\begin{subequations}
\label{eq:LLG}
\begin{align}
\dot{\ve{m}} &= \frac{1}{2}\left(\ve{\omega}_m\times\ve{m}+\ve{\omega}_n\times\ve{n}\right) + \ve{\tau}_m^\text{GD} + \ve{\tau}_m^\text{SP} + \ve{\tau}_m^\text{STT} \, ,\\
\dot{\ve{n}} &= \frac{1}{2}\left(\ve{\omega}_m\times\ve{n}+\ve{\omega}_n\times\ve{m}\right) + \ve{\tau}_n^\text{GD} + \ve{\tau}_n^\text{SP} + \ve{\tau}_n^\text{STT} \, .
\end{align}
\end{subequations}
In the LLG equation (\ref{eq:LLG}), the Gilbert damping torques are 
\begin{subequations}
	\label{eq:LLG_GD_torques}
	\begin{align}
		\ve{\tau}_m^\text{GD} &= \alpha_0\left(\ve{m}\times\dot{\ve{m}}+\ve{n}\times\dot{\ve{n}}\right)\, , \\ 
		\ve{\tau}_n^\text{GD} &= \alpha_0\left(\ve{m}\times\dot{\ve{n}}+\ve{n}\times\dot{\ve{m}}\right) \, ,
	\end{align}
\end{subequations}		
the interfacial SP torques are 
\begin{subequations}
\label{eq:LLG_SP_torques}
\begin{align}
\ve{\tau}_m^\text{SP} &= \alpha' d_\text{AF} \delta(y) \left(\ve{m}\times\dot{\ve{m}}+\ve{n}\times\dot{\ve{n}}\right) \, , \\ 
\ve{\tau}_n^\text{SP} &= \alpha' d_\text{AF} \delta(y) \left(\ve{m}\times\dot{\ve{n}}+\ve{n}\times\dot{\ve{m}}\right) \, ,
\end{align}
\end{subequations}
and the STTs are
\begin{subequations}
	\label{eq:LLG_STT_torques}
\begin{align}
\ve{\tau}_m^\text{STT} &= \frac{\alpha'}{\hbar}d_\text{AF}\delta(y) \Big[\ve{m}\nonumber\times\left(\ve{m}\times\ve{\mu}_s^\text{N}(y,t)\right) \\
&\hspace{2.15cm}+ \ve{n}\times\left(\ve{n}\times\ve{\mu}_s^\text{N}(y,t)\right)\Big]\, , \\
\nonumber\ve{\tau}_n^\text{STT} &= \frac{\alpha'}{\hbar}d_\text{AF}\delta(y) \Big[\ve{m}\times\left(\ve{n}\times\ve{\mu}_s^\text{N}(y,t)\right) \\
&\hspace{2.15cm}+ \ve{n}\times\left(\ve{m}\times\ve{\mu}_s^\text{N}(y,t)\right)\Big]\, .
\end{align}
\end{subequations}
Here, we have introduced the SP-induced enhanced damping parameter $\alpha'=\hbar\gamma g_\perp/(4\pi L A d_\text{AF})$, and $\alpha_0$ is the intrinsic Gilbert damping parameter. $A$ is the AF$|$N interface area. The STTs depend on the spin accumulation $\ve{\mu}_s^\text{N}$ in the normal metal.

The frequencies $\ve{\omega}_{m,n}$ corresponding to the effective fields are $\ve{\omega}_m=-(\gamma/L) \cdot\delta f/\delta\ve{m}$ and $\ve{\omega}_n=-(\gamma/L) \cdot \delta f/\delta\ve{n}$, where $\gamma$ is the gyromagnetic ratio and $f$ is the free energy density,
\begin{align}
\nonumber f = \frac{L}{\gamma} \Big[ &\omega_E \left(\bf{m}^2 - \bf{n}^2 \right) -2 \omega_{H} m_x \\
- &\omega_{\parallel} \left(m_x^2 + n_x^2 -(\lambda_m\nabla\ve{m})^2
-(\lambda_n\nabla\ve{n})^2\right)\Big] \,  .
\label{eq:FreeEnergy}
\end{align}
Here, $\omega_E$ is the exchange frequency, $\omega_\parallel$ is the easy-axis anisotropy frequency, $\omega_H$ is the frequency that describes the external magnetic field along the easy axis, and $\lambda_{m,n}$ is the exchange length for $\ve{m}$ and $\ve{n}$. 

\begin{figure}[h!]
\centering
\begin{tikzpicture}
\node[above right] (img) at (0,0) {\includegraphics[width=0.8\linewidth]{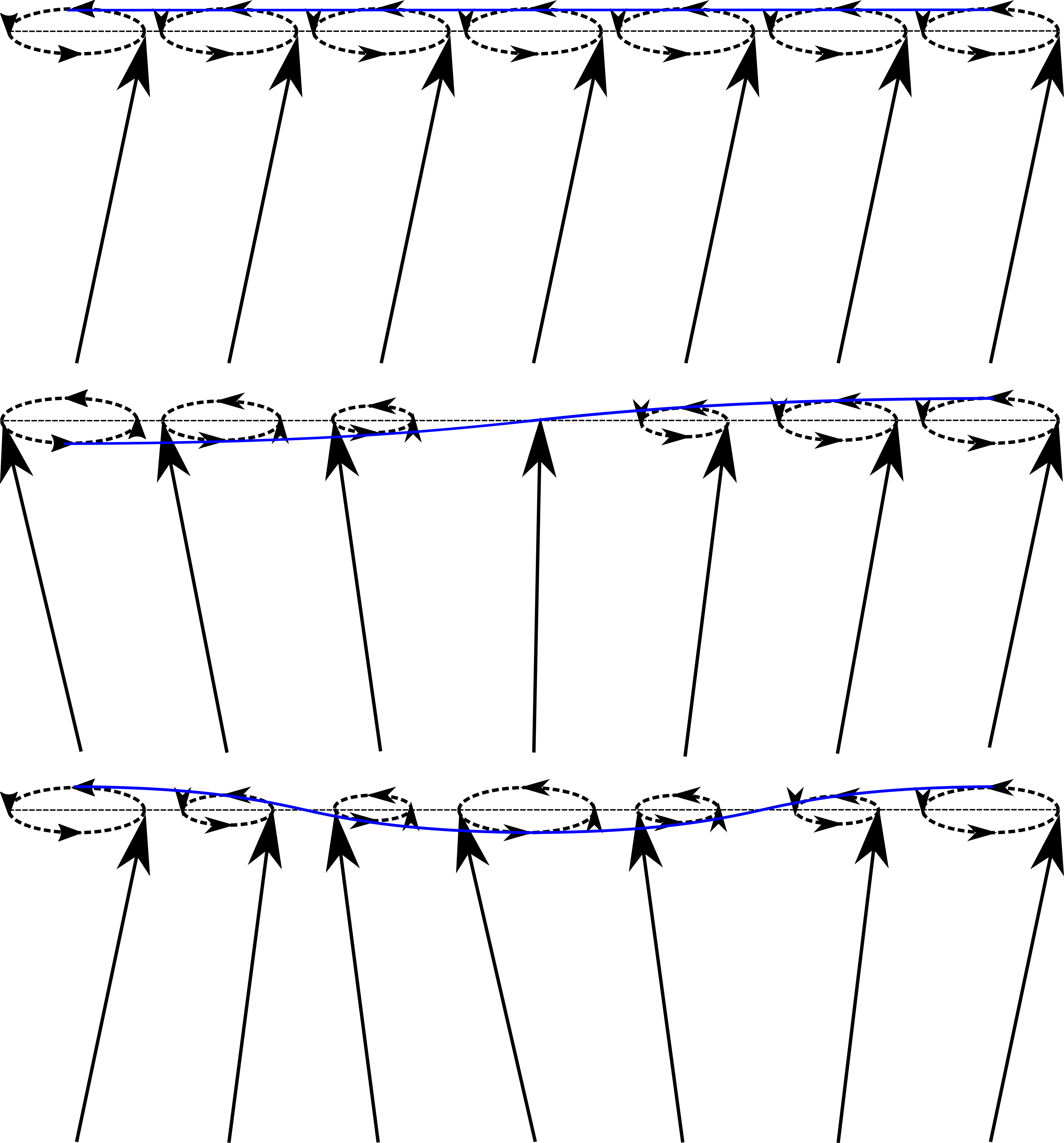}};
\node at (-10pt,180pt) {\textcolor{black}{\large{N=0}}};
\node at (-10pt,105pt) {\textcolor{black}{\large{N=1}}};
\node at (-10pt,30pt) {\textcolor{black}{\large{N=2}}};
\end{tikzpicture}
\caption{Standing waves of the N\'{e}el order parameter in the transversal direction in the limit when $\phi_{m,n}^{y,z}=0$ and $k_{m,n}^{y,z}d_\text{AF}=N\pi$.}
\label{fig:StandingWaves}
\end{figure}
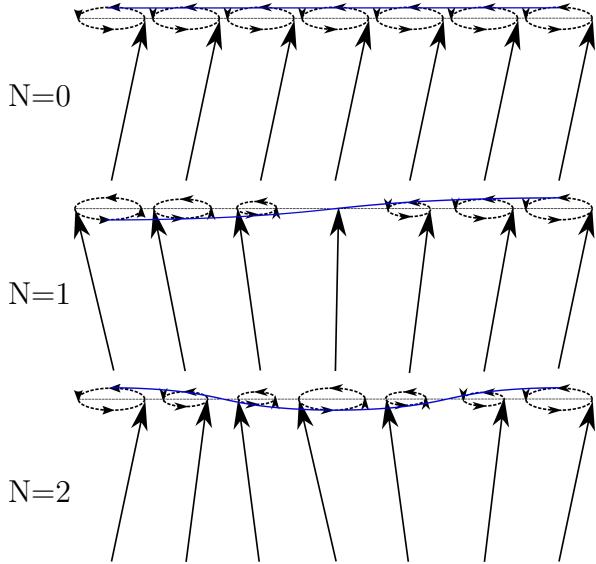

We will compute the induced dc voltages to the second order in the spin excitations. For this purpose, computing the spin excitations to the first order in their deviations from equilibrium is sufficient. For simplicity, we assume an ideal compensated antiferromagnetic insulator-metal interface. In this case, we can only excite standing waves in the transversal direction, along the interface normal. Impurities, an uneven interface, or a sufficiently high temperature can also facilitate the excitations of waves in other directions. Within our assumptions, we linearize the LLG equations and use a harmonic transversal standing-wave ansatz of the solutions:
\begin{subequations}
\label{eq:linear_ansatz}
\begin{align}
\ve{m}(y,t) &= \frac{1}{2}\left(\delta \ve{m}(y) e^{i\omega t} + \delta \ve{m}^*(y) e^{-i\omega t}\right) \, , \\
\ve{n}(y,t) &= \ve{n}_0+\frac{1}{2}\left(\delta \ve{n}(y) e^{i\omega t} + \delta \ve{n}^*(y) e^{-i\omega t}\right) \, ,
\end{align}
\end{subequations}
where $\ve{n}_0=\ve{\hat{x}}$ and
\begin{subequations}
\label{eq:linear_perturbations}
\begin{align}
\nonumber\delta\ve{m}(y) &= \delta m_y \cos[k_m^{y}(d_\text{AF}+y)+\phi_m^y]\ve{\hat{y}} \\
&+\delta m_z \cos[k_m^{z}(d_\text{AF}+y)+\phi_m^z]\ve{\hat{z}} \, ,\\
\nonumber\delta\ve{n}(y) &= \delta n_y \cos[k_n^{y}(d_\text{AF}+y)+\phi_n^y]\ve{\hat{y}} \\&+\delta n_z \cos[k_n^{z}(d_\text{AF}+y)+\phi_n^z] \ve{\hat{z}} \, .
\end{align}
\end{subequations}
Fig.\ \ref{fig:StandingWaves} illustrates the different standing waves.

\subsection{Spin accumulation}

The spin-diffusion equation determines the spatiotemporal evolution of the spin accumulation $\ve{\mu}_s^\text{N}(\ve{r},t)$ in the normal metal, 
\begin{equation}
\label{eq:SpinDiff}
\frac{\partial\ve{\mu}_s^\text{N}(\ve{r},t)}{\partial t}=\gamma_\text{N}\ve{H}_0\times\ve{\mu}_s^\text{N}+D_\text{N}\frac{\partial^2\ve{\mu}_s^\text{N}}{\partial y^2}-\frac{\ve{\mu}_s^\text{N}}{\tau_{\text{sf}}^\text{N}} \, ,
\end{equation}
where $\gamma_\text{N}$ is the gyromagnetic ratio in the normal metal, $\ve{H}_0$ is an external magnetic field, $D_\text{N}$ is the diffusion constant, and $\tau_\text{sf}^\text{N}$ is the spin-flip relaxation time.
When $\tau_\text{sf}^\text{N}$ is considerably smaller than the other time scales of the system (applied ac voltage frequency and characteristic magnetic field frequency $\omega_H=\gamma_\text{N} \abs{\ve{H}_0}$), the spin-diffusion equation can be approximated to be static. This approximation is true for metals such as Pt, which has a spin-flip relaxation time that is as low as 0.01 ps \cite{Jiao:prl:2013}. We use this simplification in our calculations, leaving us with a 1D Helmholtz equation characterized by the spin-diffusion length $\lambda_\text{sd}^\text{N} = \sqrt{D_\text{N} \tau_\text{sf}^\text{N}}$, with solutions given by hyperbolic functions.

The source of the spin accumulation in the normal metal is an ac voltage. This ac voltage leads to an ac charge current $\ve{I}_c^0(t)=I_c^0(t)\ve{\hat{x}'}$ in the metal, which generates an oscillating spin current through the spin Hall effect \cite{Dyakonov1971}. We consider a harmonic ac current with frequency $\omega$. The total spin current in the normal metal is then 
\begin{equation}
\ve{I}_s^\text{N}(y,t) = \frac{\hbar \theta_\text{SH}}{2 e} I_c^0(t)\ve{\hat{z}'} -\frac{\hbar\sigma A}{4e^2}  \frac{\partial \ve{\mu}_s^\text{N}(y,t)}{\partial y} \, .
\end{equation}
where $\theta_\text{SH}$ is the spin Hall angle and $\sigma$ is the conductivity of the normal metal. 

The boundary conditions that $\ve{\mu}_s^\text{N}$ must satisfy are that the spin current across the normal metal-vacuum interface must vanish ($\ve{I}_s^\text{N}(y=d_\text{N},t)=0$) and that the spin current across the AF$|$N interface is continuous ($\ve{I}_s^\text{N}(y=0,t)=\ve{I}_s^\text{AF}(y=0,t)$). 
The spin current in the antiferromagnetic insulator is given by contributions from SP and STTs, and it is approximated as
\begin{equation}
\ve{I}_s^\text{AF}(t) \approx \frac{g_\perp}{2\pi}\left[\hbar\left(\ve{n}\times\dot{\ve{n}}\right) + \ve{n}\times\left(\ve{n}\times\ve{\mu}_s^\text{N}\right)\right]_{y=0}
\end{equation}
to the leading order in the applied ac voltage bias, where $g_\perp$ is the transverse spin conductance. We have disregarded contributions from the imaginary part of the transverse spin conductance since it is small in most materials. 
We also consider the exchange limit ($\omega_\parallel\ll\omega_E$), which is a good approximation for many antiferromagnetic materials. 
In the exchange limit, the antiferromagnet is approximately collinear also at resonance, which means that the net magnetization is negligible. 
Any SP contributions from the magnetization will therefore be insignificant compared to the SP from the N\'{e}el order parameter. 
By solving the spin-diffusion equation (\ref{eq:SpinDiff}) with the boundary conditions, we find that  
\begin{widetext}
\begin{align}
\label{eq:mu}
\ve{\mu}_s^\text{N}(y,t) &= \mu_{s0}(t)
\frac{\sinh\left[(2y-d_\text{N})/(2\lambda_\text{sd}^\text{N})\right]}{\sinh\left[d_\text{N}/(2\lambda_\text{sd}^\text{N})\right]}\ve{\hat{z}'} 
+\frac{1}{1+\xi}\frac{\cosh\left[(y-d_\text{N})/\lambda_\text{sd}^\text{N}\right]}{\cosh\left[ d_\text{N}/\lambda_\text{sd}^\text{N}\right]}
\times\left[\hbar\left(\ve{n}\times\dot{\ve{n}}\right)-\mu_{s0}(t)\ve{n}\times(\ve{n}\times\ve{\hat{z}'})\right]_{y=0} \, ,
\end{align}
\end{widetext}
where we have introduced the dimensionless parameter 
\begin{equation}
\xi = \left[\pi\hbar\sigma A\tanh\left(d_\text{N}/\lambda_\text{sd}^\text{N}\right)\right]/(2g_\perp e^2\lambda_\text{sd}^\text{N})
\end{equation}
and a characteristic spin accumulation 
\begin{equation}
\mu_{s0}(t)=2\theta_\text{SH} e\lambda_\text{sd}^\text{N} \tanh\left(d_\text{N}/2\lambda_\text{sd}^\text{N}\right)I_c^0(t)/(A\sigma) \, .
\end{equation}

\subsection{Magnetization dynamics}

The magnetization dynamics in the antiferromagnet can be divided into two separate regions: the dynamics at the interfaces and the dynamics in the bulk. At the AF$|$N interface, the STTs $\ve{\tau}_{m,n}^\text{STT}$ drive the dynamics, and there are also dissipative SP torques ($\ve{\tau}_{m,n}^\text{SP}$). By integrating the LLG equation \eqref{eq:LLG} in a small volume around the AF$|$N interface, we find the boundary conditions for $\ve{n}$:
\begin{align}
\nonumber d_\text{AF}\alpha' &\left[\ve{n}\times\dot{\ve{n}}+\frac{1}{\hbar}\left(\ve{n}\times(\ve{n}\times\ve{\mu}_s^\text{N})\right)\right]_{y=0} \\
+\omega_\parallel &\left[\lambda_n^2\ve{n}\times\partial_y\ve{n}\right]_{y=0} = 0 \, .
\label{eq:bc_y0}
\end{align}
We assume that the other interface ($y=-d_\text{AF}$) connects to vacuum or a substrate with neither SP nor spin transfer. Subsequently, there is only a contribution to the boundary conditions from the exchange stiffness, which requires that the spatial derivative in the transversal direction vanishes.

By linearizing the boundary condition in Eq.\ \eqref{eq:bc_y0} with the ansatz in Eq. \eqref{eq:linear_ansatz}, we obtain the following constraint on the wave number $k_n^z$:
\begin{align}
\label{eq:knm_restrictions}
k_n^z d_\text{AF} \tan(k_n^z d_\text{AF}) &= i \frac{d_\text{AF}^2\alpha'\omega}{\lambda_n^2\omega_\parallel} \kappa \, ,
\end{align}
where we have introduced 
\begin{equation}
\kappa = \left[1+\frac{2e^2g_\perp\lambda_\text{sd}^\text{N}\coth(d_\text{N}/\lambda_\text{sd}^\text{N})}{A\pi\sigma\hbar}\right]^{-1} = \frac{\xi}{\xi+1} \, .
\end{equation}
When the term on the right-hand side of Eq. (\ref{eq:knm_restrictions}) is small, we can expand the solution around the roots where $k_{m,n}^{y,z}d_\text{AF}\approx N\pi$ ($N=0,1,2,\ldots$) to determine the wave numbers. 
This limit corresponds to the low-damping limit where the decay of the standing waves in the antiferromagnetic layer is negligible. 
Note that the opposite limit implies that the precessions at the interface (and thus the SP and SMR voltages) become small; therefore, this limit is of little interest.

The constraint from the boundary conditions in Eq. \eqref{eq:bc_y0} on the wave number $k_n^y$ depends on an amplitude of $\delta\ve{n}$, 
\begin{align}
&\nonumber\delta n_y \left[\lambda_n^2 k_n^y \omega_\parallel\sin(k_n^yd_\text{AF})-i d_\text{AF}\alpha'\kappa\omega\cos(k_n^yd_\text{AF})\right] \\
&=d_\text{AF}\cos\theta \alpha'\kappa\abs{\mu_{s0}}/\hbar \, .
\label{eq:ny_kny}
\end{align}
Another equation is required to find solutions for $\delta n_y$ and $k_n^y$; therefore, we must solve the LLG equations in the bulk of the antiferromagnet.

In the bulk ($-d_\text{AF}<y<0$), the LLG equation becomes a $4\times 4$ matrix equation. A non-trivial solution requires the determinant of this matrix to be zero because there is no dynamical source in the bulk. 
The dynamics enters through the boundary conditions at the interface. 
Because the determinant is independent of the precession amplitudes $\delta\ve{m}$ and $\delta\ve{n}$, we can use this condition to determine the solutions for $k_n^y$ that allow a non-trivial solution of the precession amplitudes.
The amplitude $\delta n_y$ can then be determined from Eq. \eqref{eq:ny_kny}, and the remaining amplitudes $\delta m_y$, $\delta m_z$, and $\delta n_z$ can be determined from the eigenvectors of the LLG bulk equations.

\section{Spin-transfer-torque-induced antiferromagnetic resonance}

\subsection{Frequency spectrum and susceptibilities}

In the low-damping and exchange limits, the resonance frequencies of the $N$-node standing wave mode are $\abs{\omega_\pm^{(N)}}=\omega_0^{(N)}\pm\abs{\omega_H}$, where
\begin{align}
\omega_0^{(N)} = \sqrt{\omega_0^2+\omega_\parallel\left(\frac{N\pi \lambda_{n}}{d_\text{AF}}\right)^2} \, ,
\end{align}
and $\omega_0\approx\sqrt{2\omega_E\omega_\parallel}$ is the gap frequency. 
The solutions of Eq. \eqref{eq:ny_kny} around the $k_{n}^{y,z} d_\text{AF} = N\pi$ roots are approximately complex Lorentzians,
\begin{equation}
\label{eq:nYNpm}
\delta n_{y,\pm}^{(N)}(\omega) \approx \frac{(-1)^{N+1}\alpha'^{(N)}\kappa\cos\theta \abs{\mu_{s0}}\sqrt{\omega_E}/\left(2\hbar\sqrt{2\omega_\parallel}\right)}{\abs{\omega}-\abs{\omega_\pm^{(N)}}+i \Delta\omega_\pm^{(N)}/2} \, ,
\end{equation}
where we have introduced the linewidth
\begin{equation}
\Delta\omega_\pm^{(N)} = 2\left(\alpha_0+\alpha'^{(N)}\kappa\right)\abs{\omega_\pm^{(N)}}\sqrt{\frac{\omega_E}{2\omega_\parallel}} \, .
\label{eq:linewidth}
\end{equation}
We have also introduced the effective SP damping parameter $\alpha'^{(N)}$ for the $N$-mode spin wave, where $\alpha'^{(N=0)}=\alpha'$ and $\alpha'^{(N\neq 0)}=2\alpha'$, which is analogous to the result for ferromagnetic spin waves \cite{Kapelrud:prl2013}.

The Lorentzian approximation in Eqs. \eqref{eq:nYNpm} and \eqref{eq:linewidth} is valid to the lowest order in $\alpha_0$ and $\alpha'$ under the assumption that $\lambda_{m,n}/d_\text{AF}\ll 1$.
If the antiferromagnet is so thin that the thickness becomes comparable to the exchange length, then the gap between the resonance frequencies of the macrospin mode and the higher-order standing wave modes approaches the exchange frequency $\omega_E$. 
This is the upper bound of the resonance frequency; thus, in the limit $d_\text{AF}\sim\lambda_{m,n}$, we can only excite the macrospin mode. 
Because we also want to study the higher-order standing waves, we only consider the limit where $d_\text{AF}\gg\lambda_{m,n}$.

Note that even though the linewidth is enhanced by a factor of $\sqrt{\omega_E/\omega_\parallel}$, the maximum amplitude of the precessions is not suppressed by the inverse of this factor. 
This differs from the case where the source of the dynamics is a magnetic field, where the amplitudes are suppressed by a factor of $\sqrt{\omega_\parallel/\omega_E}$.
For the spin-transfer-driven case, the only suppression arises from the high resonance frequencies of the antiferromagnet, and this suppression is also present in the magnetic-field-driven case in addition to the $\sqrt{\omega_\parallel/\omega_E}$ factor.

The $z$-component of the N\'{e}el order parameter is related to the $y$-component by
\begin{align}
\label{eq:nZNpm}
\cos\left(k_n^z d_\text{AF}\right)\delta n_{z,\pm}^{(N)} = \mp \text{sign}\left(\omega_H\right) i \cos\left(k_n^y d_\text{AF}\right)\delta n_{y,\pm}^{(N)} \, .
\end{align}
This is obtained through the eigenvectors of the bulk LLG equations in Eq. \eqref{eq:LLG} in the region where $-d_\text{AF}<y<0$. 
The magnetization $\delta\ve{m}$ also has a circular polarization for uniaxial antiferromagnets, and its amplitude is suppressed by a factor of $\propto\sqrt{\omega_\parallel/\omega_E}$ compared to the N\'{e}el order parameter. 
This suppression factor justifies our discarding of the contributions from the magnetization in the antiferromagnet to the spin accumulation in the metal. 
We now assume that the separation between the resonance frequencies $\abs{\omega_\pm^{(N)}}$ is considerably greater than the linewidths $\abs{\Delta\omega_\pm^{(N)}}$ and that the real part of the eigenfrequency is much greater than the imaginary part. The frequency spectrum for, e.g., $\delta n_y(\omega)$, can then be approximated by a sum of the complex Lorentzians in Eq.\ \eqref{eq:nYNpm}:
\begin{equation}
\delta n_y(\omega) = \sum_{N=0}^{\infty}\sum_{i=\pm}\delta n_{y,i}^{(N)}(\omega) \, ,
\end{equation}
and similarly for the other amplitudes.

\subsection{Spin Hall magnetoresistance and spin pumping dc voltages}
We can now use our solutions of the spin accumulation in Eq. \eqref{eq:mu} and precession amplitudes in Eqs. \eqref{eq:nYNpm} and \eqref{eq:nZNpm} to determine the total charge current resulting from the applied voltage and interaction with the antiferromagnet driven at resonance. 
This charge current density is \cite{Chen:prb:2013}
\begin{equation}
\ve{j}_c(y,t) = \frac{I_c^0(t)}{A}\ve{\hat{x}'}+\frac{\theta_\text{SH}\sigma}{2e}\ve{\hat{y}}\times\frac{\partial\ve{\mu}_s^\text{N}(y,t)}{\partial y} \, .
\end{equation}
By averaging over the normal metal, $\overline{\ve{j}_c}(t) = d_\text{N}^{-1}\int_0^{d_\text{N}}\ve{j}_c(y,t)\diff{y}$, we find that the contributions to the $x'$-direction are
\begin{equation}
\overline{\ve{j}_c}(t)\cdot\ve{\hat{x}'} = j_{c,x'}^\text{SMR}(t)+j_{c,x'}^\text{SP}(t),
\end{equation}
where
\begin{align}
\label{eq:jSMR}
j_{c,x'}^\text{SMR}(t) &= \frac{I_c^0(t)}{A}\left[1-\frac{\Delta\rho_0}{\rho}-\frac{\Delta\rho_\text{S}}{\rho}(1-n_{z'}^2)\right]_{y=0} \, , \\
\label{eq:jSP}
j_{c,x'}^\text{SP}(t) &=-\frac{\theta_\text{SH}\hbar\sigma}{2d_\text{N} e}\eta\left[\left(\ve{n}\times\dot{\ve{n}}\right)_{z'}\right]_{y=0} \, .
\end{align}
Here, we have introduced 
\begin{align}
\Delta\rho_0 &= -\rho \theta_\text{SH}^2\frac{2\lambda_\text{sd}^\text{N}}{d_\text{N}}\tanh\left(\frac{d_\text{N}}{2\lambda_\text{sd}^\text{N}}\right) \, , \\
\eta &= \frac{1}{1+\xi} \tanh\left(\frac{d_\text{N}}{2\lambda_\text{sd}^\text{N}}\right) \tanh\left(\frac{d_\text{N}}{\lambda_\text{sd}^\text{N}}\right) \, ,
\end{align}
and the SMR $\Delta\rho_\text{S}=-\eta\Delta\rho_0/2$. $\rho=1/\sigma$ is the resistivity of the normal metal. 
The contributions from both the SMR and the SP induce a dc component in the resulting ISHE voltage in the normal metal. 
Assuming that $I_c^0(t) = I_c^0\cos(\omega t)$, we find that
\begin{align}
\label{eq:jSMR_dc}
\langle j_{c,x'}^\text{SMR}(t)\rangle_t
= \frac{\Delta \rho_\text{S} I_c^0}{2\rho A} \sin2\theta\Real \left[\delta n_z \cos(k_n^z d_\text{AF})\right] \, .
\end{align}
To find the dc contributions from SP, we study the dc component of $\langle (\ve{n}\times\dot{\ve{n}})_{z'}\rangle_t$, and we compute that
\begin{align}
\nonumber \langle (\ve{n}\times\dot{\ve{n}})_{z'} \rangle_t = -\omega \Imag \Big[( & \delta n_y\cos(k_n^y d_\text{AF}))^* \\
\times & \delta n_z\cos(k_n^z d_\text{AF})\Big]\sin\theta.
\end{align}

Let us now compare the results for the dc components of $j_{c,x'}^\text{SMR}$ and $j_{c,x'}^\text{SP}$ in Eqs. \eqref{eq:jSMR} and \eqref{eq:jSP} to the ferromagnetic case \cite{Chiba:pra:2014}.  We observe that the results are exactly the same when $\ve{n}\leftrightarrow\hat{\ve{M}}$ and $G_r\rightarrow 2G_r$, where $\hat{\ve{M}}$ is the magnetization unit vector in the ferromagnet and $G_r$ is the real transverse spin conductance in Ref. \onlinecite{Chiba:pra:2014}. 

Experiments measuring the SMR in NiO$|$Pt heterostructures indicate that the SMR is negative for antiferromagnets \cite{Hou:prl:2017,Hoogeboom2017,Baldrati2017,Fischer2017}. 
Because the only key difference between the antiferromagnetic case and the ferromagnetic case is that the N\'{e}el order parameter, not the magnetization, causes the SMR, the negative sign must be due to some property of the N\'{e}el order parameter.
This is in agreement with the reasoning in Ref.\ \onlinecite{Hoogeboom2017}, where the negative SMR is explained by the coupling of the N\'{e}el order parameter to the magnetic field. 
They typically couple perpendicularly to each other, whereas for ferromagnets, the magnetization couples along the magnetic field. 
The perpendicular coupling gives rise to a $\pi/2$ phase shift relative to the ferromagnetic case and a negative sign in the measured SMR.

If we consider the case in which the susceptibility of the N\'{e}el order parameter is of the same order of magnitude as the susceptibility of the magnetization in a ferromagnet, then the SMR and SP voltages in an antiferromagnet should be comparable to those in a ferromagnet. Eq. \eqref{eq:nYNpm} shows that the susceptibility scales with the inverse of the resonance frequency. The susceptibility of the N\'{e}el order parameter therefore becomes comparable to that of the magnetization in a ferromagnet when the system is driven close to the spin-flop transition, where the resonance frequency is small \cite{Johansen:prb:2017}. 

Inserting our solutions of the frequency-dependent amplitudes in Eqs. \eqref{eq:nYNpm} and \eqref{eq:nZNpm}, the dc voltages as a function of applied ac voltage frequency become approximately
\begin{widetext}
\begin{subequations}
\label{eq:VDC}
\begin{align}
V_\text{dc}^\text{SMR}\left(\omega\right) &= \text{sign}\left(\omega_H\right) K \sin 2\theta\cos\theta \left(\frac{I_c^0}{A}\right)^2 \sum_{N=0}^{\infty}\sum_{i=\pm} \frac{\alpha'^{(N)}}{2\left(\alpha_0+\kappa\alpha'^{(N)}\right)} \frac{i\cdot L_i^{(N)}\left(\omega\right)}{\abs{\omega_i^{(N)}}} \, , \\
V_\text{dc}^\text{SP}\left(\omega\right) &= -\text{sign}\left(\omega_H\right)\kappa K \sin 2\theta\cos\theta \left(\frac{I_c^0}{A}\right)^2 \sum_{N=0}^{\infty}\sum_{i=\pm}\left[\frac{\alpha'^{(N)}}{2\left(\alpha_0+\kappa\alpha'^{(N)}\right)}\right]^2 \frac{i\cdot L_i^{(N)}\left(\omega\right)}{\abs{\omega_i^{(N)}}} \, ,
\end{align}
\end{subequations}
\end{widetext}
where $V_\text{dc}^\text{SMR/SP}=l\rho \langle j_{c,x}^\text{SMR/SP}(t)\rangle_t$ and $l$ is the length of the bilayer in the direction of the applied voltage. 
We have also introduced the symmetric Lorentzian 
\begin{equation}
L_i^{(N)}(\omega) = \frac{\left(\Delta\omega_i^{(N)}/2\right)^2}{\left(\abs{\omega}-\abs{\omega_i^{(N)}}\right)^2+\left(\Delta\omega_i^{(N)}/2\right)^2} \,
\end{equation}
and the constant
\begin{equation}
K = \frac{l \kappa \eta \theta_\text{SH}^3 e \left(\lambda_\text{sd}^\text{N}\right)^2}{\hbar d_\text{N} \sigma^2} \tanh^2\left[\frac{d_\text{N}}{2\lambda_\text{sd}^\text{N}}\right] \, .
\end{equation}

In our model, the SMR and SP voltages as functions of frequency are described via symmetric Lorentzians. However, we have not included the contributions from the Oersted field to the dynamics. The charge current causes an oscillating magnetic field that leads to an antisymmetric Lorentzian component \cite{Chiba:pra:2014}. One therefore needs to filter out the antisymmetric component before comparing experimental data with our model. We did not take  the Oersted field in the free energy into account since the susceptibility associated with this magnetic field is a factor of $\sim \sqrt{\omega_\parallel/\omega_E}$ smaller than the susceptibility associated with the spin accumulation \cite{Sluka2017}.
Moreover, because the Oersted field is approximately uniform in a sufficiently thin antiferromagnetic film, it can only couple to the $N=0$ mode. The symmetric Lorentzian can therefore be expected to be the dominant component of the signal for most antiferromagnetic materials and should be the only component for the $N\neq 0$ modes.

\begin{table}[h!]
	\centering
		\caption{Material parameters for MnF$_2$.} 
	\begin{tabular}{l l l} \hline \hline
	 $\omega_E$ (s$^{-1}$) \cite{Ross:TUMunchen:2013}& $\omega_{\parallel}$ (s$^{-1}$) \cite{Ross:TUMunchen:2013}& $L$ (A/m) \cite{Kotthaus:prl:1972}\\ \hline
	9.3$\cdot 10^{12}$ & 1.5 $\cdot 10^{11}$ & 47 862 \\ \hline \hline
	\end{tabular}
	\label{tab:MnF2}
\end{table}
\begin{table}[h!]
	\centering
		\caption{Material parameters for Pt.} 
	\begin{tabular}{l l l} \hline \hline
	 $\theta_\text{SH}$ \cite{Obstbaum:prb:2014} & $\lambda_\text{sd}^\text{Pt}$ (nm) \cite{Meyer:apl:2014}  & $\sigma$ ([$\Omega$m]$^{-1}$) \cite{Liu2011} \\ \hline
	0.12 & 1.5 & 5$\cdot 10^6$ \\ \hline \hline
	\end{tabular}
	\label{tab:Pt}
\end{table}

Next, we will compute the dc voltages for MnF$_2$ using the parameters in Table \ref{tab:MnF2} and Table \ref{tab:Pt}. Direct measurements of some of the parameters are lacking. We therefore use these missing material parameters from similar systems. 
We use the Gilbert damping of NiO, $\alpha_0=2.1\cdot10^{-4}$ \cite{Kampfrath2011}, and the typical transverse conductance for ferromagnet-normal metal systems, $g_\perp/A \sim 3\cdot10^{18}$ m$^{-2}$ (ferromagnet$|$Pt \cite{Yoshino2011}). 

The exchange length is $\propto a \sqrt{\omega_E/\omega_\parallel}$, where $a$ is the typical lattice spacing. We define $\lambda_{n}=a z\sqrt{\omega_E/\omega_\parallel}$, where $z$ is the number of nearest-neighbor sites on each sublattice. Because MnF$_2$ has a tetragonal crystal structure and therefore two lattice constants \cite{Dormann1977}, we use the average of the two lattice constants for our characteristic length $a$. MnF$_2$ has eight nearest neighbors. This results in an estimated exchange length of $26$ nm. We note, however, that the real value might differ by as much as an order of magnitude from this value. The exchange length should therefore be estimated by measuring the separation between  the resonance peaks. 

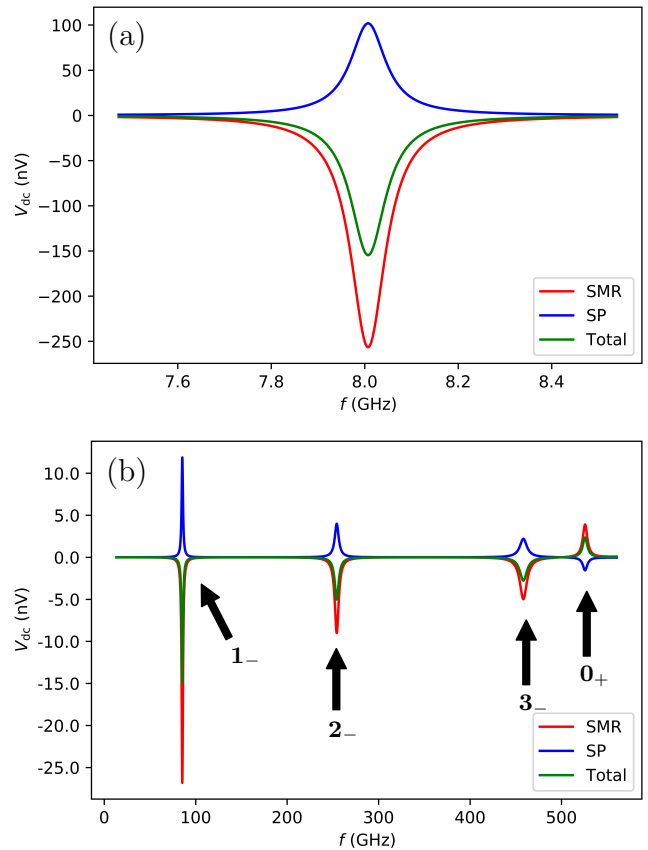
\begin{figure}[h!]
\centering
\begin{tikzpicture}
\node[above right] (img) at (0,165pt) {\includegraphics[width=\linewidth]{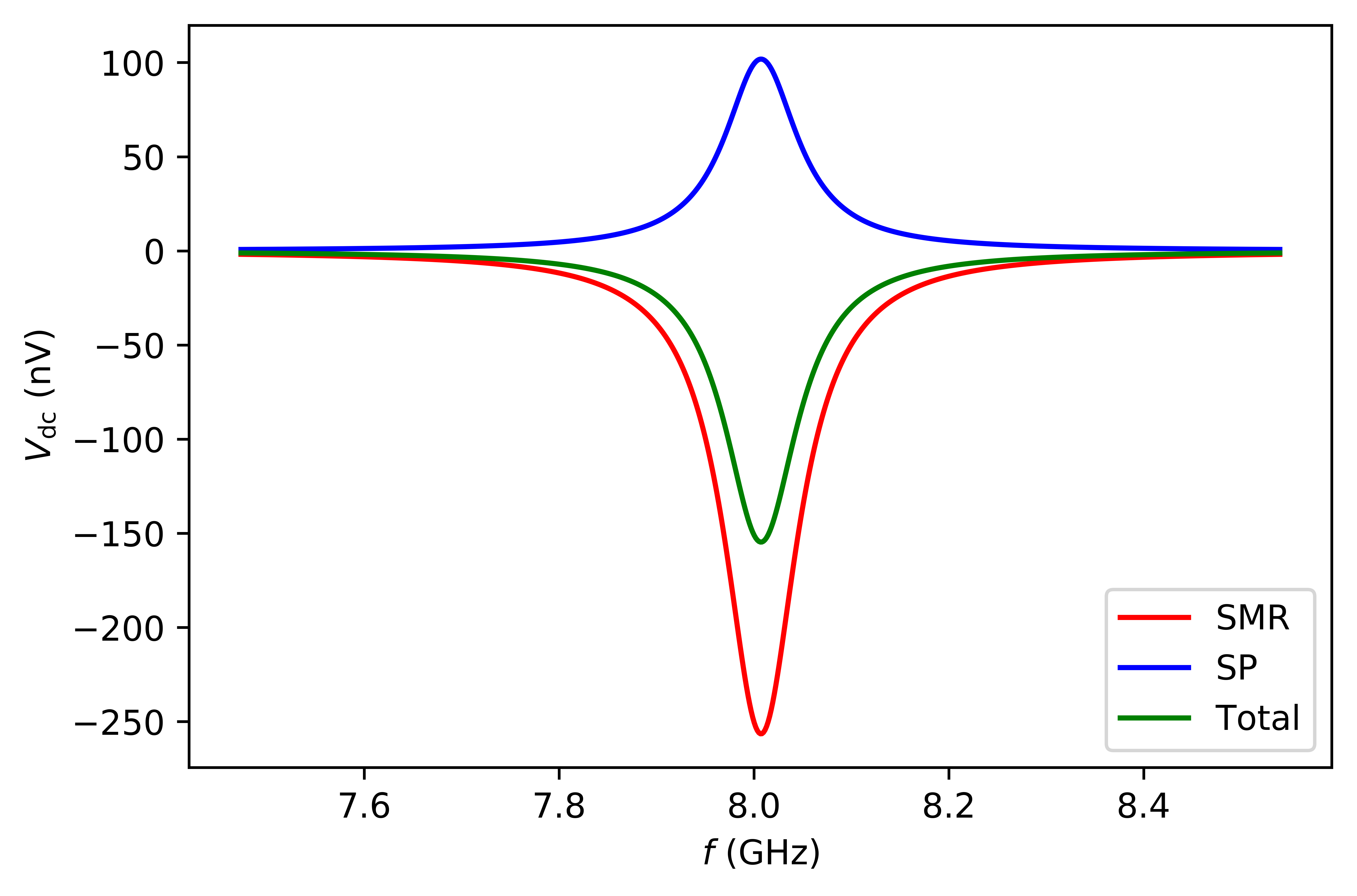}};
\node[above right] (img) at (0,0) {\includegraphics[width=\linewidth]{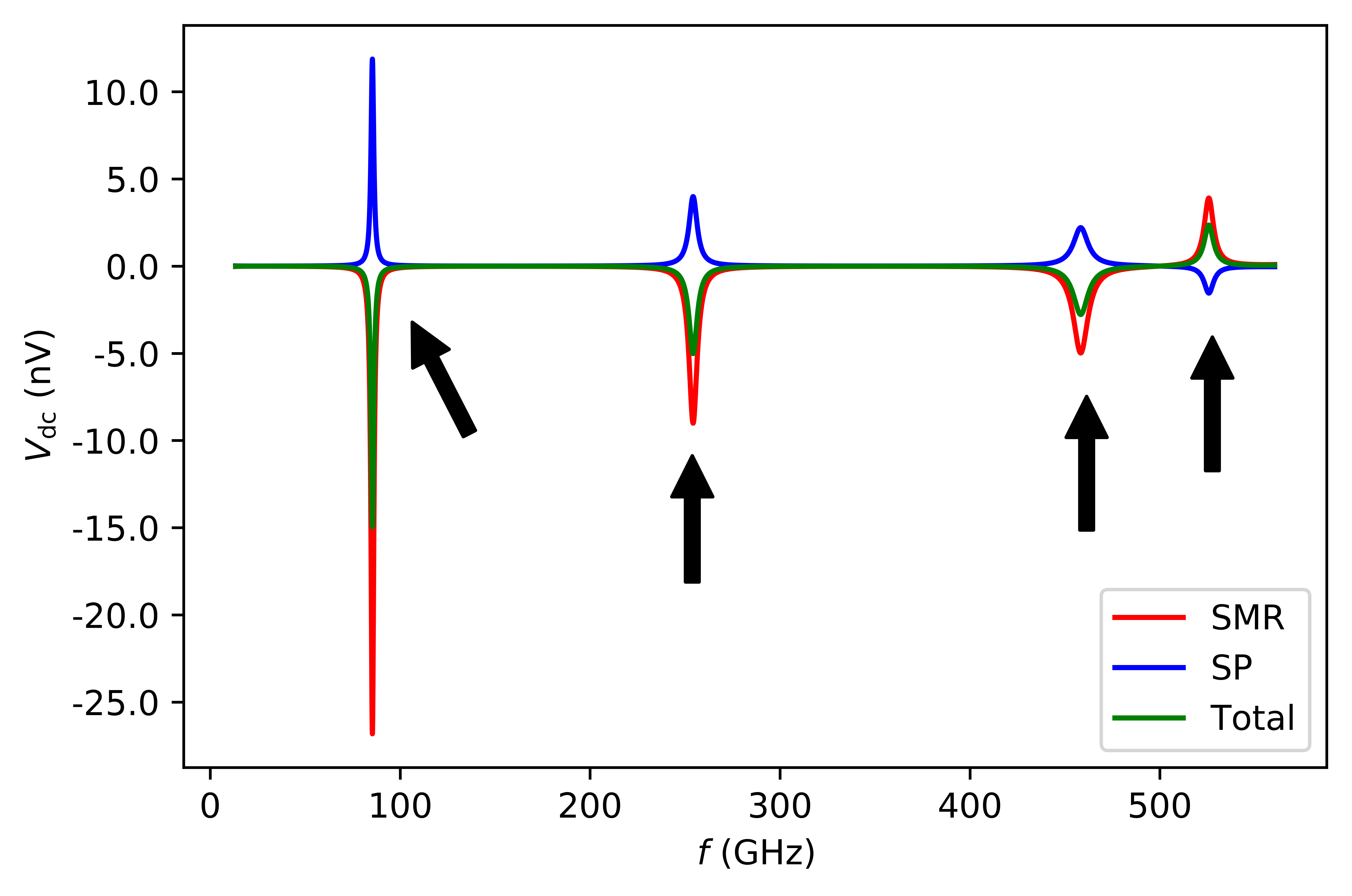}};
\node at (50pt,315pt) {\textcolor{black}{\large{(a)}}};
\node at (50pt,150pt) {\textcolor{black}{\large{(b)}}};
\node at (95pt,80pt) {\textcolor{black}{$\mathbf{1_-}$}};
\node at (132pt,52pt) {\textcolor{black}{$\mathbf{2_-}$}};
\node at (204pt,62pt) {\textcolor{black}{$\mathbf{3_-}$}};
\node at (227pt,73pt) {\textcolor{black}{$\mathbf{0_+}$}};
\end{tikzpicture}
\caption{Resonance spectrum for a MnF$_2$ film of thickness $d_\text{AF}=100$\ nm in an external magnetic field $\omega_H=0.97\omega_0$. The dc voltages for the low-energy macrospin mode ($0_-$) are shown in (a), while some of the higher-energy left-handed $N_-$ modes and the right-handed macrospin mode ($0_+$) are shown in (b).}
\label{fig:SMR_SP_Voltages}
\end{figure}

The results obtained using these parameters are presented in Figure \ref{fig:SMR_SP_Voltages}, where we have used $l=100$\ $\mu$m, $d_\text{N}=2\lambda_\text{sd}^\text{Pt}$, $I_c^0/A = 10^{10}$ A/m$^2$, and $\theta=35^\text{o}$. The SMR and SP dc voltages always have opposite signs, whereas their frequency dependence is exactly the same. The partial cancellation leads to a smaller net signal. The contributions from SMR and SP cannot be distinguished from one another in this bilayer system.  We also observe that the SP voltage is always smaller than the SMR voltage. For a given direction of the external magnetic field, here assuming that $\omega_H>0$, the signs of the dc signals depend solely on whether the precessions are right handed ($\omega_\text{res}>0$, $+$ mode) or left handed ($\omega_\text{res}<0$, $-$ mode). 
The dc voltages resulting from the higher-energy modes are not particularly large for our choice of parameters. 
This is primarily due to the high resonance frequencies of these modes. 
The standing waves will be easier to detect in materials with a lower gap frequency $\omega_0$ than MnF$_2$ or in materials with a shorter exchange length, which leads to a lower resonance frequency for these modes. 
Examples of antiferromagnets with a low gap frequency are RbMnF$_3$, which has a gap of $\omega_0/2\pi=9$ GHz \cite{Ince1966}, and GdFe$_3$(BO$_3$)$_4$, which has a gap of $\omega_0/2\pi=29$ GHz \cite{Pankrats2004}. 
For comparison, the gap frequency of MnF$_2$ is $\omega_0/2\pi=267$ GHz. 
Because experimental data for the exchange lengths in antiferromagnets are lacking, it is difficult to suggest material candidates for this category. 
We expect, however, that the exchange length is proportional to $\sqrt{\omega_E/\omega_\parallel}$; thus, the exchange length is small in materials where the easy-axis anisotropy is significant compared to the exchange energy. 
The disadvantage of this is that the exchange energy is generally large in antiferromagnets; thus, the gap frequency and spin-flop fields will typically also be large for these materials. 
An example of this case is FeF$_2$, which has $\omega_\parallel = 0.37 \omega_E$ \cite{Ohlmann:prb:1961}. 
The gap frequency for this material is as large as 1.41 THz, and the spin-flop field is 50.4 T. 
Spin-transfer antiferromagnetic resonance experiments with this material will therefore be very challenging.
As an alternative to finding materials with lower resonance frequencies, one can apply a stronger voltage to enhance the signals because the measured dc voltages are quadratic in the applied ac voltage.

In the next section, we will discuss a trilayer system that allows separating the SMR and SP voltages. Separating these voltages yields more information about the system, such as the ratio between the intrinsic damping to the SP-enhanced damping.

\section{Separation of spin Hall magnetoresistance and spin-pumping voltages}

\subsection{N$|$AF$|$N system}

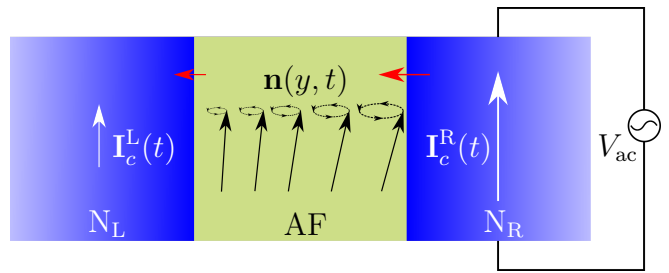
\begin{figure}[h!]
\centering
\begin{tikzpicture}
\node[above right] (img) at (0,0) {\includegraphics[width=\linewidth]{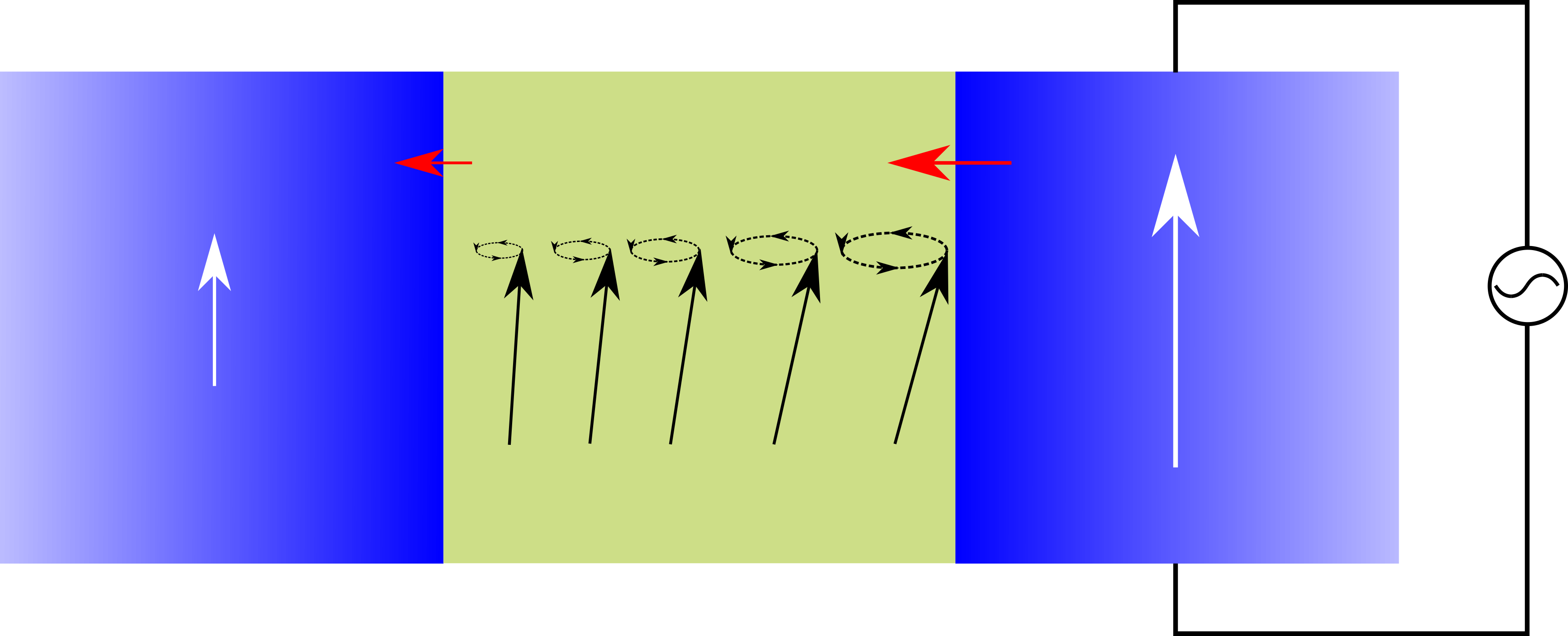}};
\node at (233pt,50pt) {\textcolor{black}{\large{$V_\text{ac}$}}};
\node at (173pt,50pt) {\textcolor{white}{\large{$\ve{I}_c^\text{R}(t)$}}};
\node at (190pt,21pt) {\textcolor{white}{\large{N$_\text{R}$}}};
\node at (115pt,21pt) {\textcolor{black}{\large{AF}}};
\node at (40pt,21pt) {\textcolor{white}{\large{N$_\text{L}$}}};
\node at (54pt,50pt) {\textcolor{white}{\large{$\ve{I}_c^\text{L}(t)$}}};
\node at (115pt,75pt) {\textcolor{black}{\large{$\ve{n}(y,t)$}}};
\end{tikzpicture}
\caption{An antiferromagnet sandwiched between two normal metals. Magnetization dynamics in the antiferromagnet is induced by applying an ac voltage with a constant amplitude on the normal metal to the right, which leads to a spin current into the antiferromagnet. A spin current is then pumped into the left normal metal, which induces a charge current through the inverse spin Hall effect.}
\label{fig:NAFN}
\end{figure}

We now extend and generalize our considerations to an antiferromagnetic insulator sandwiched between two normal metals, as illustrated in Fig. \ref{fig:NAFN}. We apply an ac voltage with a constant amplitude to an active normal metal as in the previous sections. Additionally, we use a passive normal metal to detect the SP contributions from the antiferromagnet. Because the passive normal metal does not exhibit any SMR dc voltage to the leading order in the applied voltage source, we measure the SMR and SP voltages independently. 

Taking advantage of the symmetry of our system, the spin accumulation in the passive normal metal can readily be obtained from Eq. \eqref{eq:mu}. Because the only source for the spin accumulation is SP from the antiferromagnet, we find the spin accumulation to be
\begin{align}
\nonumber \ve{\mu}_s^\mathrm{N_L}(y,t) = 
&\frac{\hbar}{1+\xi}\left[\ve{n}\times\dot{\ve{n}}\right]_{y=-d_\text{AF}} \left(\cosh\left[ d_\mathrm{N_L}/\lambda_\text{sd}^\mathrm{N_L}\right]\right)^{-1}\\
&\times\cosh\left[(y+d_\mathrm{N_L}+d_\text{AF})/\lambda_\text{sd}^\mathrm{N_L}\right] \, .
\label{eq:mu_NL}
\end{align}
Consequently, the average charge current in the passive normal metal along the $x'$-direction becomes
\begin{align}
j_{c,x'}^\text{L}(t)=\overline{\ve{j}_{c}^\text{L}}(t)\cdot\ve{\hat{x}'} &=\frac{\theta_\text{SH}\hbar\sigma}{2d_\mathrm{N_L} e}\eta\left[\left(\ve{n}\times\dot{\ve{n}}\right)_z\right]_{y=-d_\text{AF}} \, .
\end{align}
If we, for simplicity, assume no decay of the spin current in the antiferromagnet and let the properties and dimensions of both normal metals be identical, then we can observe from Eq. \eqref{eq:jSP} that $\langle j_{c,x'}^\text{L}(t)\rangle_t=-\langle j_{c,x'}^\text{SP}(t)\rangle_t$. In other words, we can indirectly measure the SMR dc voltage by measuring the ratio of the dc voltage in the passive normal metal $\mathrm{N_L}$ relative to the dc voltage in the active normal metal $\mathrm{N_R}$. This indirect measurement of the SMR voltage assumes that the decay of the pumped spin current is insignificant. We will now determine in what region this approximation holds.

\subsection{Spin-current decay}

The non-zero spin current across the left AF interface implies that the boundary conditions at $y=-d_\text{AF}$ must be extended to include SP and STTs:
\begin{align}
\nonumber d_\text{AF}\alpha' &\left[\ve{n}\times\dot{\ve{n}}+\frac{1}{\hbar}\left(\ve{n}\times(\ve{n}\times\ve{\mu}_s^\text{N})\right)\right]_{y=-d_\text{AF}} \\
-\omega_\parallel &\left[\lambda_n^2\ve{n}\times\partial_y\ve{n}\right]_{y=-d_\text{AF}} = 0 \, .
\label{eq:bc_LLG_NAFN}
\end{align}
The boundary conditions at $y=0$ remain unchanged and are given by Eq. \eqref{eq:bc_y0}. 
As a result of the new boundary conditions at $y=-d_\text{AF}$, the solutions for the phases $\phi_{m,n}^{y,z}$ in our linear response ansatz in Eq. \eqref{eq:linear_perturbations} are no longer zero as they were in the bilayer system. 
The phases will now have a finite correction in $\alpha'$.
We can rewrite our boundary conditions at $y=0,-d_\text{AF}$ to the following constraints on the wave numbers and phases:
\begin{subequations}
\label{eq:k_phi}
\begin{align}
\label{eq:phi_n}
k_n^{y,z}d_\text{AF}\tan\phi_n^{y,z} &= -i\frac{d_\text{AF}^2\alpha' \omega}{\lambda_n^2\omega_\parallel}\kappa \, , \\
k_n^{z}d_\text{AF}\tan\left(k_n^{z}d_\text{AF}+\phi_n^{z}\right) &= i\frac{d_\text{AF}^2\alpha' \omega}{\lambda_n^2\omega_\parallel}\kappa \, .
\end{align}
\end{subequations}
We can decouple the above equations to obtain constraints that are only dependent on the wave number $k_n^z$,
\begin{align}
&\nonumber\tan\left(k_n^{z}d_\text{AF}\right)\left[k_n^{z}d_\text{AF}+\frac{1}{k_n^{z}d_\text{AF}}\left(\frac{d_\text{AF}^2\alpha' \omega}{\lambda_n^2\omega_\parallel}\kappa\right)^2\right] \\
\approx \ &k_n^{z}d_\text{AF}\tan\left(k_n^{z}d_\text{AF}\right)=2 i \frac{d_\text{AF}^2\alpha' \omega}{\lambda_n^2\omega_\parallel}\kappa \, .
\label{eq:knz_trilayer}
\end{align}
This constraint is similar to the constraint for the AF$|$N bilayer in Eq. \eqref{eq:knm_restrictions} to the lowest order in $\alpha'$, except that $\alpha'\rightarrow2\alpha'$. The doubling of the damping due to SP is because we now pump spins across two interfaces rather than one interface.
The last constraint on $k_n^y$ is equivalent to Eq. \eqref{eq:ny_kny}, where we now also have to take  the non-zero phase $\phi_n^{y}$ into account; thus, the boundary condition becomes
\begin{align}
\nonumber d_\text{AF}\cos\theta &\alpha'\kappa\abs{\mu_{s0}}/\hbar= \delta n_y \big[\lambda_n^2 k_n^y \omega_\parallel\sin(k_n^yd_\text{AF}+\phi_n^y) \\
-i d_\text{AF} &\alpha'\kappa\omega\cos(k_n^yd_\text{AF}+\phi_n^y)\big] \, .
\label{eq:ny_kny_trilayer}
\end{align}

The decay of the spin current in the antiferromagnetic insulator is related to the imaginary components of $d_\text{AF}k_{n}^{y,z}$. 
At resonance and to the lowest order in $\alpha'$ and $\alpha_0$, we find these to be
\begin{subequations}
\label{eq:ImagComponentsMacro}
\begin{align}
\label{eq:ImagComponentsMacroz}
\left|\Imag{\left(d_\text{AF}k_{n,N=0}^{z}\right)}\right| &= \sqrt{\frac{d_\text{AF}^2\kappa\alpha'\omega}{\lambda_{n}^2\omega_\parallel}} \, , \\
\label{eq:ImagComponentsMacroy}
\left|\Imag{\left(d_\text{AF}k_{n,N=0}^{y}\right)}\right| &= \sqrt{\frac{d_\text{AF}^2\left(\kappa\alpha'+\alpha_0\right)\omega}{\lambda_{n}^2\omega_\parallel}} \, ,
\end{align}
\end{subequations}
for the macrospin mode and 
\begin{subequations}
\label{eq:ImagComponentsN}
\begin{align}
\label{eq:ImagComponentsNz}
\left|\Imag{\left(d_\text{AF}k_{n,N>0}^{z}\right)}\right| &= \frac{2 d_\text{AF}^2 \kappa\alpha' \omega}{\lambda_{n}^2 N\pi \omega_\parallel} \, \\
\label{eq:ImagComponentsNy}
\left|\Imag{\left(d_\text{AF}k_{n,N>0}^{y}\right)}\right| &= \frac{d_\text{AF}^2 \left(2\kappa\alpha' +\alpha_0\right)\omega}{\lambda_{n}^2 N\pi \omega_\parallel} \, ,
\end{align}
\end{subequations}
for the standing-wave modes, respectively. 

Let us now study how the imaginary components in Eqs. \eqref{eq:ImagComponentsMacro} and \eqref{eq:ImagComponentsN} scale with $d_\text{AF}$ and the resonance frequency $\omega$.
Since the SP-induced damping $\alpha' \propto 1/d_\text{AF}$, $\Imag{\left(d_\text{AF}k_{n,N}^{z}\right)}$ scales as $\propto \left(d_\text{AF}\omega\right)^\zeta$, where $\zeta=1/2$ for the macrospin mode and $\zeta=1$ for the standing waves ($N>0$). 
In the limit where $\alpha_0\ll\alpha'$, when the bulk damping is small compared to the interface damping, $\Imag{(d_\text{AF}k_{n,N}^{y})}$ scales as $\Imag{(d_\text{AF}k_{n,N}^{z})}$. 
However, when $\alpha_0$ becomes large compared to $\alpha'$, the bulk damping dominates, and $\Imag{(d_\text{AF}k_{n,N}^{y})}$ scales as $\propto \left(d_\text{AF}^2\omega\right)^\zeta$. 
We can then observe, as expected, that the spin current decays faster as a function of $d_\text{AF}$ for thicker films where the bulk damping dominates. 
Based on these scaling relations, we observe that we can minimize the decay of the spin current, thereby keeping the magnitude of the SP at the two interfaces similar to each other, by (i) keeping the antiferromagnetic layer sufficiently thin and (ii) reducing the resonance frequency by driving the system close to the spin-flop transition.

Assuming that the SP is dominated by the dynamics of $\ve{n}$, which is a good assumption for most collinear antiferromagnets, the transmission of the pumped spin current through the antiferromagnetic layer can be defined as
\begin{equation}
\label{eq:TSP}
T_\text{SP}^{(N)} = \frac{\langle\left(\ve{n}\times\dot{\ve{n}}\right)_{y=-d_\text{AF}}\rangle_t}{\langle\left(\ve{n}\times\dot{\ve{n}}\right)_{y=0}\rangle_t} \, .
\end{equation}
This describes the ratio of the SP at the passive interface relative to the active interface where we excite the dynamics by injecting a spin current. 
When this ratio is close to unity, it is a good assumption that we are in the low-decay regime, and the pumped spin current across the two interfaces will be approximately the same. 
In the low-decay regime, the SMR and SP dc voltages can be separated by measuring the dc voltage in both normal metals independently.

\begin{figure}[h!]
\centering
{\includegraphics[width=\linewidth]{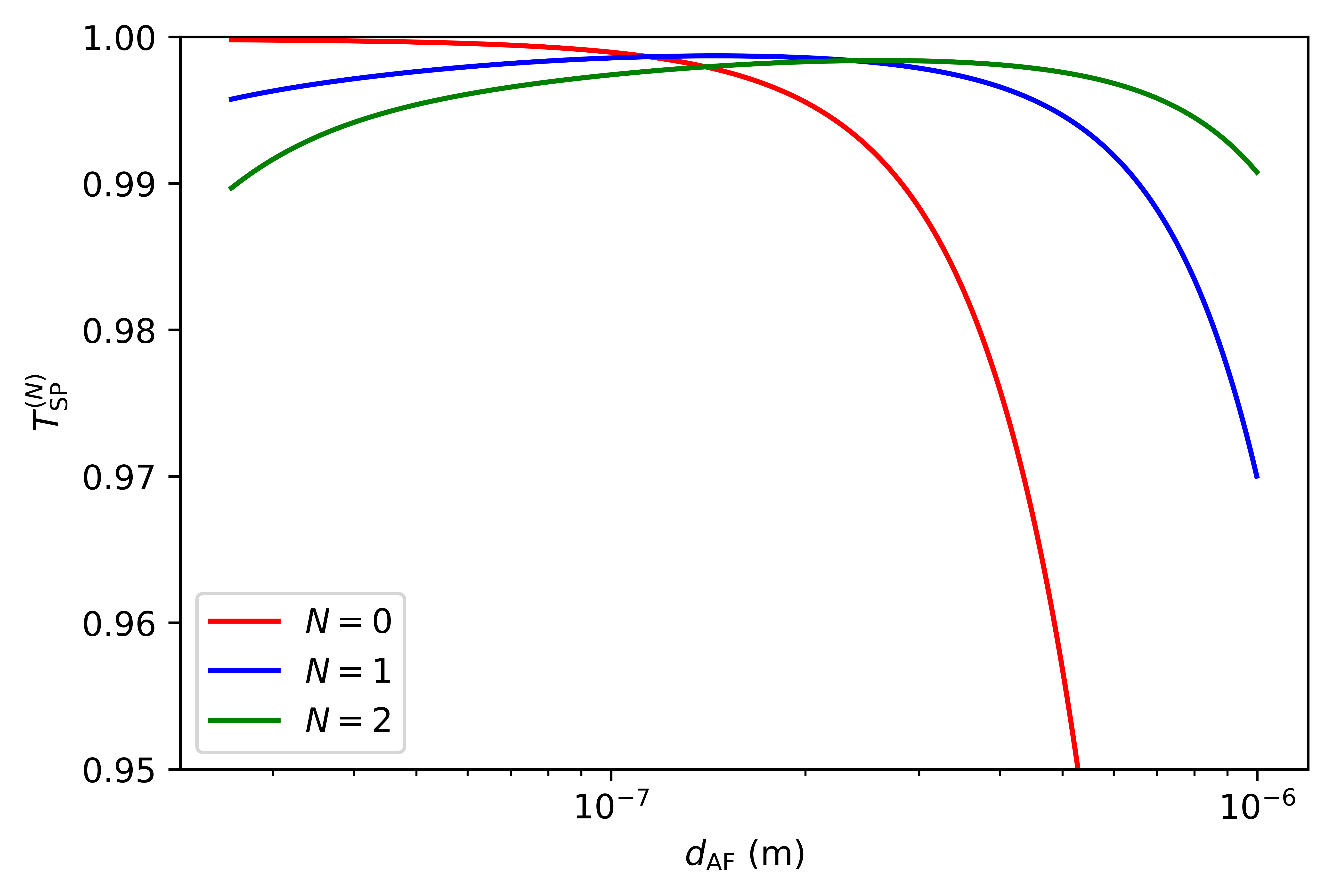}}
\caption{Ratio of the spin pumping at the passive AF$|$N interface relative to the spin pumping at the active AF$|$N interface as a function of $d_\text{AF}$ for three of the low-energy left-handed $N_-$ modes at $\omega_H = 0.9\omega_0$.}
\label{fig:TSP}
\end{figure}

We plot the transmission of the pumped spin current as a function of $d_\text{AF}$ in Fig. \ref{fig:TSP} for the three lowest energy modes. 
As shown, the pumped spin current of the modes with the lowest energy exhibits the fastest decay with increasing thickness of the antiferromagnetic layer. 
This result is as expected from the scaling behaviors in Eqs. \eqref{eq:ImagComponentsMacro} and \eqref{eq:ImagComponentsN}, as $d_\text{AF}^2\alpha'\omega\kappa/(\lambda_n^2\omega_\parallel)$ is a small dimensionless number for the choice of parameters that we have previously considered. 
If $d_\text{AF}^2\alpha'\omega\kappa/(\lambda_n^2\omega_\parallel)$ is of order unity or larger, then we are in the large-damping limit, and we can therefore expect the decay of the spin wave amplitudes to become significant.
We can also observe that the $N>0$ modes decay when $d_\text{AF}\sim\lambda_{n}$ in addition to large values of $d_\text{AF}$, unlike the macrospin mode. 
This result is due to the high resonance frequencies of the standing waves in this limit. 

It is now interesting to determine at what thickness the bulk damping starts to become more important compared to the interface damping. 
For our choice of parameters, we have that $\alpha'=\alpha_0$ at $d_\text{AF}=440$ nm. 
As shown in Fig. \ref{fig:TSP}, below this value, the transmission is close to unity, and the pumped spin current quickly decays as we move into the region where the bulk damping dominates. 
This does not mean that the decay can be neglected in the limit where the interface damping dominates, as this can also be significant if $d_\text{AF}$ is very large compared to the exchange length $\lambda_n$ or if the resonance frequency is high. 
However, this result clearly indicates that the transmission of the spin current will quickly decay with increasing $d_\text{AF}$ as the bulk damping starts to dominate, which is in agreement with our scaling analysis.

Based on our results, one can observe that $T_\text{SP}^{(N)}\approx 1$ is a good approximation when $\Imag{\left(d_\text{AF} k_n^{y,z}\right)} \ll 1$. 
We can then utilize the analytical expressions for these imaginary components in Eqs. \eqref{eq:ImagComponentsMacro} and \eqref{eq:ImagComponentsN} to evaluate whether we are in a low-decay regime, where the SMR and SP dc voltages can be separated.

\section{Conclusions}
We have studied STT-induced antiferromagnetic resonance in bilayers that consist of an antiferromagnetic insulator and a normal metal and in a metal-antiferromagnetic insulator-metal trilayer. 
We consider excitations of the uniform mode and of the transverse standing waves. The dc voltages have contributions from the SMR and SP, similar to ferromagnetic systems. 
In the antiferromagnetic system, the dynamics of the N\'{e}el order parameter causes these effects. A challenge in an antiferromagnetic system is the weak signals due to the low susceptibility of the N\'{e}el order parameter. We demonstrate how the signals are enhanced by driving the system close to the spin-flop transition, where the resonance frequency is lower. In trilayer systems, the contributions due to SP and SMR can be separated when the antiferromagnetic layer is thin.

\section*{Acknowledgments}
This work was supported by the Research Council of Norway through its Centres of Excellence funding scheme, project number 262633 “QuSpin” and Grant No. 239926 "Super Insulator Spintronics", as well as the European Research Council via Advanced Grant no. 669442 "Insulatronics". \newline

\textit{Note} -- During the completion of this work, a recent independent study of spin-transfer antiferromagnetic resonance was reported \cite{Sluka2017}. Ref. \onlinecite{Sluka2017} computes the spin accumulation and frequency dependence of the conductivity in the normal metal for the macrospin mode. 
A main point and difference in our work is that we consider a magnetic field that importantly reduces the frequency and enhances the output signal. This facilitates experimental detection in an experimentally feasible frequency range.
Additionally, we study the excitation of standing spin waves and a trilayer system that can be utilized to separate the output signals resulting from SP and SMR.

\appendix
\bibliography{bibliography}
\end{document}